\magnification 1200
  \centerline {\bf Marketing Percolation}

  \bigskip
J. Goldenberg$^1$, B. Libai$^2$, S. Solomon$^3$, N. Jan$^4$, and D. Stauffer$^5$

  \bigskip
  $^1$ School of Business and Administration, Mt. Scopus, Hebrew University,
  Jerusalem 91905, Israel

  $^2$ Davidson Faculty of Industrial Engineering and Management,
  Technion, Haifa 32000, Israel.

  $^3$ Racah Institute of Physics, Givat Ram, Hebrew University, Jerusalem
  91904, Israel

  $^4$ Physics Department, St Francis Xavier University,
     Antigonish, Nova Scotia, Canada B2G 2W5

  $^5$ Institute for Theoretical Physics, Cologne University,
       D-50923 K\"oln, Germany

  e-mail: sorin@vms.huji.ac.il

  \bigskip
  Abstract:
  A percolation model is presented, with computer simulations for illustrations,
  to show how the sales of a new product may penetrate the consumer market.
  We review the traditional approach in the marketing literature, which is
  based on differential or difference equations similar to the logistic
  equation (Bass 1969). This mean field approach is contrasted with the
  discrete percolation on a lattice, with simulations of "social percolation"
  (Solomon et al 2000) in two to five dimensions giving power laws instead
  of exponential growth, and strong fluctuations right at the percolation
  threshold.

  \bigskip
  \bigskip

  1. {\bf Introduction.}

  If the amount of activity in an academic area reflects its importance,
  then research on the diffusion of innovations, with over 4,000 diffusion
  publications since 1940, is one of the most important areas in the social
  sciences. ``No other field of behavioral science research represents more
  effort by more scholars
in more disciplines in more nations'' [1]. Marketing's considerable share of the
output in this research stream reflects not only the importance of new products,
but also the role of diffusion research in helping managers to better plan their
entry strategy, target the right consumer and anticipate demand so as to have an
  efficient and effective promotion, production and distribution strategy.
  (We use here the terminology of marketing theory
  and call them diffusion models, whereas the physics of diffusion is quite a
  different process.)
Diffusion as marketing experts define it is the development (increase) of
sales over time (not spatial again), it is viewed as analogous to epidemics with
increasing number of ill people. So is the new product: increasing number of
adopters is in essence the diffusion process.
  The growth of new products is a complex process which typically consists
  of a large body of consumers interacting with each other over a long
  period of time.

  Distressingly, often only aggregate data on adoption
  (i.e. the sum of all previous sales etc.)
  is available to researchers for analysis, as is generally the case with
  market level diffusion models [2,3].
  (Aggregate data means that the sales are measured once in a quarter or a year,
   without any attention to spatial distribution, and no attention to the
   individual buyer.)
  Even when collecting data at the individual level, diffusion research surveys
  consist of correlated data gathered in one ``snap shot'' survey of consumers,
  a methodology that amounts to freezing the diffusion process, making the
  continuous time-dependent process timeless [1].

  Hence, it is not surprising that much of the theoretical base to the
  diffusion of innovations is grounded on repeatedly analyzed small number
  of data sets, in which researchers could actually follow the diffusion
  process within small social systems, such as the cases of the diffusion
  of hybrid corn among farmers in Iowa [4], antibiotics among US physicians [5]   or family planning in Korean villages [6]. While the impressive
  contribution of these studies is evident, new tools should be considered
  to analyze the fast changing and complex environment of new product
  growth.

  The small set of available individual based data poses another
  research dilemma: The small number of cases can not offer us an over-view
  of how collective behavior emerges from changes in individual
  characteristics. The span of individual level parameters is too small to
  allow for developing an explanation of their relations to the diffusion
  parameters or to predict them from the diffusion parameters.

  Thus, the modeling of the diffusion of new products lies between two extremes.

  Aggregate, or market level, diffusion models, such as the Bass model [7],
  an equation similar to what physicists call the logistic equation or Verhulst
factor, are based on market level data and assume a large degree of homogeneity
in the population of adopters.
Basically, the diffusion of innovations models primary
  premise is based on the assumption that communication between individuals,
  is central to the new product's growth.

  One of the advantages of diffusion models is that they provide a relatively
easy and parsimonious analytic way to look at the whole market and interpret its
  behavior, yet, still based on rich and empirically based theory. Another
  advantage is that very often the market level is also the level managers
  will be mostly interested in. Finally, aggregate models can be estimated
  with market level data such as number of adoptions in a given year or
  average price, which are relatively easy to get.

  This simplicity is also associated with some critique on the aggregate
  approach to diffusion. One shortcoming is that the models make strong
  and simplifying assumptions on the behavior of individuals, for example
  the lack of heterogeneity among adopters. Also, the ability to test the
  assumptions these models make with very limited data at the aggregate
  level can be questioned [8].

  Individual level models, on the other hand, acknowledge differences
  between consumers (e.g., difference in utility among potential adopters
  and their affect on adoption). Generally they follow economic theories
  (e.g. [9]) and assume that individuals maximize some
  personal objective function such as utility of the product, and may
  update their beliefs as more information arrives at the market. Thus,
  individual level models can be viewed as more behaviorally based than
  aggregate models.

  Aiming at explaining aggregate adoptions in the market level,
  restrictions on the heterogeneity in behavior among individuals are
  sometimes introduced, and  individual levels models are aggregated to
  provide an explicit diffusion function at the market level (e.g.,
  [10]). Yet, the use of market level data to
  calibrate individual level models is still not very common, partly
  because the very limited aggregate level data do not really allow
  individual level testing, as the case in the traditional diffusion
  models.

  Our study synthesizes individual and aggregate level modeling in a way
  which may help to overcome some of the outlined barriers. We follow
  diffusion theory and its emphasis on the communication behavior as a
  driver of new product growth, and generate a variety of possible dynamics
  to explore their influence on the aggregate level. Percolation
  enables us to perform sensitivity analysis and examine the effect of
  changes in the parameters in the individual level on the aggregate
  level, and thus overcome some of the limitations that follow the use of
  few data points at the aggregate level.

  \bigskip
  2: {\bf Diffusion Models: A Background}

  New products (in particular really new products) undergo a diffusion
  process: From an initial stage (in which there are zero buyers)
  individuals start to adopt the innovation and buy the product until the
  relevant market completely adopts it.
  Diffusion models try to explain and predict diffusion rates as a
  function of type of innovation, communication channels, nature of the
  social systems etc. Despite the large number of factors the models are
  parsimonious. The history of diffusion research in marketing is briefly
  presented below:

  \medskip
  a) 1969: The Bass model

  The modeling of the aggregate penetration of new products in the
  marketing literature generally follows the Bass model [7]. The model
  follows Rogers' diffusion of innovations theory of 1962 [1] which emphasizes
  the role of communication methods: external influence (e.g.,
  advertising, mass media) and internal influence (e.g. WOM = Word Of
   Mouth),  as driving the product adoption pattern. Thus, an individual's
  probability of adopting a new product at time $t$ (given that s/he had not
  adopted yet) depends in the Bass model linearly on two forces: a force
  which is not related to previous adopters and is represented by the
  parameter of external influence (traditionally denoted as $p$), and a
  force that is related to the number of previous adopters, the parameter
  of internal influence (denoted as $q$).  The hazard model that describes
  the conditional probability of adoption at time $t$ is:
  $$ f(t)/[1-F(t)] = p+qF(t)  \eqno (1)$$
  where $f(t)$ is the probability of adoption at time $t$ and $F(t)$ describes
  the cumulative probability of adoption.
  Generally, $p$ represents the effect of external influences, i.e.,
  influence not related to the number of previous adopters, such as
  advertising. $q$ represents the effect of internal influence, coming from
  previous adopters.

  In the marketing practice eq.(1) is used in the form of eq.(2) in which
  $n(t)$ represents the buyers (or adopters) within a specified time interval
  and $N(t)$ is the cumulative number of buyers in a market of $M$ possible
  buyers:
  $$    n(t) = [p + q(N(t)/M)][M-N(t)] \eqno (2)$$

  The Bass model has four main properties:
i) It is the most dominant and popular. ii) It fits well many data.
iii) After enough data points it is used in practice to forecast sales.
iv) However its relevance to a real consumer behavior is questioned in
several papers.
Its significance (at least to the marketing people) lies
also in the fact that the two main parameters can represent internal effects
(due to previous adopting population) and external effects (not related to
previous adopting population). In that it follows the diffusion of innovations
theory, one of the well known theories of social sciences, that attributes the
adoption rate of innovations to communication processes such as Word of Mouth
from previous adopters (an internal effect) and mass media influence (an
external effect).

  \medskip

  b) 1978-79: Extensions of the basic Bass models

  Modifications to increase the precision of the model in various cases
  were suggested. As an example consider a new class of flexible diffusion
  models, which allow non-symmetric patterns, heterogeneous adopters
  population etc. Those modifications were motivated by the need for
  better fit to real life data. A typical model from this generation is:
  $$    n(t) = [a + b(N(t)/c(M(t))^{1+d}][cM(t) - N(t)]^{1+e} \eqno (3) $$
  where M is the market potential and A, b, c, d, e are estimated from the data.

  \medskip
  c) 80's and 90's : more growth models

  During the 80's data on product penetration and diffusion were
  accumulated and diverse patterns were observed leading to suggestion of
  models with different penetration curves. Since growth modeling is an
  important occupation in a lot of fields, the marketing literature
  benefits from other fields' achievements. But the main occupation
  consisted of tailoring a Bass-type model to a specific segment of
  innovation adoption. For example eq. 4 below was found to fit well
  adoptions of durables in the agricultural context.
  $$  n(t) = b[N(t)/M(t)]\; [\ln M/n(t)]                   \eqno (4)$$
  In many cases these models do not relate to diffusion theory, rather
  they offer smoothing of a noisy data better then other regression
  technique. Furthermore, their relevance to marketing is sometimes
  criticized [8] because they have little direct marketing application.

  \bigskip
  {\bf 3. Shortcomings in this approach}

  Indeed, this research stream produced many extensions incorporating
  assumption regarding issues such as the effect of marketing mix,
  competition, repeat purchase and technological substitution (see, for
  example, reviews [3,8,11]). The prediction ability was reported to be
  satisfactory for various practical implications. However, it seems that
  this aggregate modeling approach reaches its limits. For instance, how
  can the coefficients of the smoothing function be interpreted in the
  individual level?  This does not come straightforward from the Bass
  model equation, where $p$ and $q$ are part of the linear combination that
  governs the hazard rate.

  Aggregate diffusion models make very simplifying assumptions that assume
  homogeneity in the communication behavior of adopters. However, while
  concern regarding this issue has been expressed throughout the diffusion
  literature (e.g. [12]), because of the nature of
  the very aggregate data available to researchers, limited options were
  available to those who wanted to examine these assumptions, and their
  implications.

  In this paper we demonstrate how a microscopic presentation (more precisely
  percolation modeling) can be used to link market level models to
  individual level behavior. Further, it will allow us to examine the
  effect of heterogeneity in the communication behavior of adopters on the
  aggregate adoption level that are typically analyzed in aggregate
  diffusion models such as the Bass model and its extensions [3].
  In short, we replace the prevailing mean field theory
  by a more microscopic statistical approach,
  taking into account fluctuations and spatial correlations.

  \bigskip
  {\bf 4. The percolation representation of product diffusion}

  Our technique is at once simple, direct and very powerful: represent in
  the computer the individual buyers, products and sales as well as the
  information transfer, and the changes in their current individual
  status.

Each site $i$ of a large lattice is occupied with a random number $p_i$ between
  zero and one, representing the customer's quality expectation. The quality
  of a new product is called $Q$, and potential customers buy it only if this
  quality is above their expectations: $Q > p_i$ [13]. This
  standard percolation model [14,15] has a
  critical percolation threshold $p_c$ such that for $Q > p_c$ an infinite
  cluster of neighboring buyers can be formed, while for $Q < p_c$ all clusters
  of buyers are finite. There is a formal equivalence between this picture and
  the marketing phenomena: far below a certain quality level the
  product does not sell at all, while far above that density of buyers, the
  product reaches most of its potential market.
  The percolation literature [14,15] contains much information about the spatial
  geometry of clusters which could be used for market modelling.

  As long as the $p_i$ do not change, the cluster structures for different $Q$
  are correlated.
  This suggests that one can use the recorded dynamics of one sweep in
  order to predict the behavior of the subsequent ones, or in general in
  order to characterize the cluster structure of the market.
  This line of thought is natural in the context of microscopic simulation
  but is quite novel in marketing.

  Of course in reality even a product which "makes it"
  may produce losses if the producer over estimates its market share and
  keeps producing after this is exhausted. On the other hand, the
  fluctuations (which the percolation model predicts) may discourage a
  producer and lead him to discontinue the production (flop) even in
  conditions in which the product could "make it".

  In addition to the basic capability to express detailed spatio-temporal
  knowledge on the market structure and behavior the model above
  introduces  significant conceptual departures from the main features and
  assumptions  of the Bass model.

  \medskip
  a) In the Bass model, the fluctuations around the Bass formula are
  assigned to measurements errors or to repeat purchases (especially close
  to the peak) in the microscopic simulation the
  fluctuations are the result of the random irregularities in the
  connectivity between various parts of the system.
  In fact fluctuations in the sales rate can appear even if one excludes
  the  possibility of repeated purchases. In particular one can identify
  strongly connected clusters within which the sales front advances fast
  separated by regions poor in potential buyers which correspond to
  stagnation or slowing down in the sales.

  \medskip
b) We could generalize this site percolation picture to a site-bond percolation
  model, where connections between neighboring buyers are formed only with some
  bond probability; and then these bond probabilities increase with time if
  neighboring sites have bought a product. Then,
  if one allows for the existence of successive product waves (annual issues
  of the car models, movies in a series like star wars, pink panther,
  etc.) one obtains smoother curves and larger clusters than  in the initial
  wave. This is due to the emergence of "battered paths": herds of buyers
  with regularly coordinated coherent response to the product.
  One can say that once the connections between the buyers clusters are
  established, they are less effective in slowing the propagation of the
  product sales front.

  \medskip
  c) The functional form of the Bass curve is affected too: rather than an
  exponential increase during the entire pre-saturation region, one
  reaches a linear (or in general power)  region of sales increase once a
  clear propagation front is formed. Indeed, the surface of a
  $d$-dimensional ball increases as the radius to the power $d-1$.

  \medskip
  d) The products which take off are the ones which happen to be planted
  in a  cluster rich in potential buyers. If the cluster is small, the
  product sales will halt upon reaching the cluster boundary. If the
  cluster is large, the sales  will be much higher.

  \medskip
  e) Note that the usual Bass model is just the exact solution of the
  model in the extreme case where the "neighbors" which link to each site
  are chosen randomly on the lattice (then the effect of common neighbors
  is negligible as long as the finite size is negligible) and in which
  instead of having a buy or a refuse to buy at each site one has always
  a buy (possibly with a different quantity expressing the buyer
  preference).
  In that case, which ignores the discrete character of the buying event
  (and the discrete choices of the discrete buyers) one gets an
  exponential increase followed by saturation and one has a perfect
  averaging of the buying rate by the various quantities bought by the
  buyers at the buying front.

  \medskip
  f) In the case when the product quality $Q$ and the quality expectation $p_i$
  change in time [13], the adaptation of the buyers tastes and the producers
  offer (in terms of quality and price) has the effect that after a few
  waves of similar
  products the system will be roughly [17] at the boundary between the
  exponential decay and exponential increase of sales.  Moreover one
  observes a certain convergence of tastes of the buyers towards a common
  behavior or towards separate groups which have convergent behavior
  within the group and divergent between the groups (large regions which
  react to a new product in a series coherently within the group and
  disjoint among the groups).

  \medskip
  g) On top of this one can consider modeling the effects of peer
  pressure: sites which are not potential buyers becoming buyers when many
  of the neighbors bought the product. Sometimes this is not just a
  psychological effect: it is related with the utility of the product
  depending on its use by the other buyers (like in the case of fax, ps,
  pdf, word files formats).
  \medskip

  h) The above simple percolation model with time-independent $p_i$ and $Q$
  was used to produce in our figure various curves for $n(t)$, the number of
  new buyers which are neighbors to site which have already bought the product
  in the previous time interval. We start with one buyer and then let the buying
  spread over the lattice with a Leath algorithm [16]: At each time step,
  all neighbors of all previous buyers decide, once and for all, if they buy.
Thus our time steps are microsteps in the sense of Huang [17], for one spread of
  one product through the lattice, and not macrosteps in the sense of
  Solomon et al [13] referring to repeated attempts with different parameters
  like $Q$. We see in fig.1 single examples below, at and above the percolation
  threshold (no infinite cluster, one fractal infinite cluster, and one
  compact infinite cluster, respectively).
  In reality the time resolution may be less fine than in these simulations;
  this could be
  taken into account by binning together several consecutive time steps and thus
  reducing the short-time fluctuations without changing the long-time trends.
  Fig.2 shows averages over many samples at the percolation threshold; then the
  fluctuations vanish. Fig.3 shows such results also for higher dimensions $d$
  where the initial sales grow roughly as $t^{d-1}$, approaching for $d
  \rightarrow \infty$ the exponential growth of the traditional theories like
  eq.1. (For figures 2 and 3 we used the fully dynamic model [13,17] where
  $q$ and the $p_i$ self-organize towards the percolation threshold $p_c$ in
  steps of 0.001; for fixed $q = p_c$ and fixed $p_i$ as for fig.1,
  the slopes are smaller and depend on whether we average over all clusters or
  only over the ``infinite'' clusters.)
  For comparison, Fig.4 shows two examples of real markets, for automobiles
  and for LCD color television sets in the 20th century, indicating,
  respectively, an exponential increase (or power law with a large exponent
  like 5) and a linear growth. The first example may be better described by
  a Bass-type theory, the second better by two-dimensional percolation as in
  Fig. 3.

 i) As is customary for phase transitions in physics,
the percolation transition implies that even if the
probability distribution of the $p_i$ across the lattice is
totally uniform, one ends up with localized clusters and sub-clusters
of all scales including macroscopic inhomogeneity leading to macroscopic
sales rate fluctuations. The fractal
clustered character of the market and the bottlenecks
are not detectable by the usual polling techniques.
A uniform customer distribution
leads at the percolation threshold to un-passable barriers and to
(almost-) extinction of sales. Similarly, a minor increase of temperature
can make the water boiling, without any change in the intermolecular forces.
\bigskip
{\bf 5. Review of Cluster Geometry}

We summarize here some well known percolation properties which could become
relevant in a future more quantitative theory of marketing along our outlines.
The geometry of percolation clusters has been studied since decades [14,15].
Their surface should not be defined as the set of empty neighbors of occupied
sites since the number of such empty neighbors is proportional to the number of
occupied sites. Instead, the fractal dimension $D$ is a widespread quantitative
measure, defined through $M \propto R^D$ for large clusters with radius $R$
and $M$ occupied sites. On two-dimensional lattices, $D$ is about 1.6, 1.9
and exactly 2 for $q$ below, at and above $p_c$. For the largest cluster at
$p_c$ one can replace the radius $R$ by the linear lattice dimension $L$ in the
above mass-radius relation. The fluctuations in the mass of the largest cluster
are at $q = p_c$ about as large as the average mass, even if $L$ goes to
infinity.

The largest cluster at the percolation threshold, also called the incipient
infinite cluster, consists mostly of dangling ends, that means of links through
which no current flows if a voltage is applied to two points of the cluster.
The current-carrying part of the backbone, varying as $L^{1.6}$, and consists
mostly of sites which can be removed without cutting the cluster into parts.
Red sites are the bottlenecks, removal of which cuts the cluster into separate
parts; their number increases only as $L^{0.75}$. The number $\ell$ of sites
which link within the incipient infinite cluster two sites at Euclidean distance
$r$ is called the chemical distance and corresponds to the time (microsteps)
needed in our model to transfer information from one customer to the other;
it varies on average as $r^{1.1}$ for large distances.

All these exponents are valid rather generally in two dimensions, not only for
nearest-neighbor connections on the square lattice. For example, we could allow
information to flow also to the four next-nearest neighbors in addition to
the four nearest neighbors. Then the value of $p_c$ would change but the
above exponents would be the same and thus universal. Only if this distance of
neighbors goes to infinity do we expect a behavior more similar to eqs.(1-4).
Thus these exponents as opposed to $p_c = 0.592746$ are rather general
quantitative predictions of percolation theory to be compared with future
high-precision data from real marketing.

  \bigskip
  {\bf 6. Summary}

  The traditional approach towards marketing theory, in the literature cited
  here, has been replaced by a percolation model which treats each customer
  individually instead of averaging over all of them. As a result strong
  fluctuations are observed, as found in real sales curves. Just as other
percolation applications, also the present one can be modified in numerous ways
  to describe better specific effects.

  We thank Z.F. Huang for discussions and
K-mart International Center of Marketing and Retailing,
Davidson Center, Alexander Goldberg Academic Lectureship Fund, German-Israeli
Foundation, NSERC of Canada and SFB 341 of Germany for partial support.

  \bigskip

{\bf References}

\parindent 0pt
1. Rogers, E. M. (1995). The Diffusion of Innovations, 4th, New York:
Free Press.

\medskip
2. Sultan, F., Farley, J. U., and Lehmann, D. R. (1990). "A Meta
Analysis of Applications of Diffusion Models," Journal of Marketing
Research, 27, 70.

\medskip
3. Mahajan, V., Muller, E., and Bass, F. M. (1990). "New Product
Diffusion Models in Marketing: A Review and Directions for
Research,"Journal of Marketing, 54, 1.

\medskip
4. Ryan B. and Gross, N.C.  (1943). "The Diffusion of Hybrid Seed
Corn in Two Iowa Communities," Rural Sociology, 8, 15.

\medskip
5. Coleman, J.S., Katz, E., and Menzel, H. (1966). Medical Innovation: A
Diffusion Study. Bobbs-Merrill, New York.

\medskip
6. Rogers E. and Kincaid, D.L. (1981). Communication Networks: A New
Paradigm for Research. New York: Free Press.

\medskip
7. Bass, F. M., (1969). "A New Product Growth Model for Consumer
Durables," Management Science, 15, 215.

\medskip
8. Parker, P. M. (1994)."Aggregate Diffusion Models in Marketing: A
Critical Review,"
International Journal of Forecasting 10, 353.

\medskip
9. Lancaster, K. (1966). "A New Approach to Consumer Theory," Journal
of Political Economy 74, 132.

\medskip
10. Chatterjee, R. and Eliashberg, J. (1990). " The Innovation
Diffusion Process in a Heterogeneous Population: A Micro Modeling
Approach," Management Science, 36, 1057.

\medskip
11. Bass, F. M., Trichy, K., V., and Jain, D. C. (1994). "Why
the Bass Model Fits Without Decision Variables," Marketing Science,
13, 203.

\medskip
12. Mahajan, V. and Peterson, R. A. (1985). Models for Innovation
Diffusion, Sage, Newbury Park, CA.

\medskip
13. Solomon, S.  and Weisbuch, G. e-print adapt-org 9909001; Solomon, S.
Weisbuch, G., de Arcangelis L., Jan, N. and Stauffer, D. Physica A 277, 239
(2000)

\medskip
14. Stauffer, D. and Aharony, A. (1994). Introduction to Percolation Theory.
Taylor and Francis, London.

\medskip
15. Sahimi, M. (1994). Applications of Percolation
Theory. Taylor and Francis, London.

\medskip
16. Evertz, H. G. (1993). J. Stat. Phys. 70, 1075.

\medskip
17. Huang, Z.F. (2000). Int. J. Mod. Phys. C 11, 287 and Eur. Phys. J. B,
in press.

\medskip
18. {\it The Electronic Market Data Book}, Electronic Industries Alliance,
Washington DC, 2000.
\bigskip

{\bf Figure Captions}

Fig.1. Examples of simulated sales curves below (line), at (b, +)
and above (x) the percolation threshold (part a). Parts b and c show the cluster
geometry for the three curves of part a, after the cluster stopped growing or
touched the upper boundary.

Fig.2. Averaged sales curves at the percolation threshold for $L \times L$
squares with $L$ = 100, 200, 500 and 1001 (from left to right): The individual
fluctuations are washed out in the average.

Fig.3. Averaged sales curves for two to five dimensions; the straight lines
give the theoretically expected slopes $d-1$ in this log-log plot.

Fig.4. Yearly sales (in thousands) in the USA of automobiles (a) and of
LCD color television sets (b), versus year [18].
\end